\documentclass[12pt]{amsart}
\usepackage[mathscr]{euscript}
\usepackage{amscd}
\usepackage{amsmath}
\theoremstyle{plain}
\newtheorem{theorem}{Theorem}[section]
\newtheorem{lemma}[theorem]{Lemma}
\newtheorem{proposition}[theorem]{Proposition}

\theoremstyle{definition}

\theoremstyle{remark}

\numberwithin{equation}{section}
\newcommand{\gb}{\overline{g}}
\newcommand{\hb}{\overline{h}}

\newcommand{\Xc}{\overline{X}}
\newcommand{\M}{\mathcal M}
\newcommand{\ep}{\epsilon}
\newcommand{\up}{\Upsilon}
\newcommand{\eh}{\hat{\epsilon}}
\newcommand{\rh}{\hat{r}}
\newcommand{\ch}{\hat{c}}
\newcommand{\gh}{\hat{g}}

\newcommand{\al}{\alpha}
\newcommand{\om}{\omega}
\newcommand{\Om}{\Omega}
\newcommand{\be}{\beta}
\newcommand{\ga}{\gamma}
\begin{document}
\title{Conformal Anomaly Of Submanifold Observables in
AdS/CFT Correspondence}

\author[R. Graham]{C. Robin Graham}
\address{
  Department of Mathematics\\
  University of Washington \\
  Box 354350\\
  Seattle, WA 98195-4350}
\email{robin@math.washington.edu}

\author[E. Witten]{Edward Witten}
\address{ School of Natural Sciences \\
          Institute for Advanced Study \\
          Princeton, NJ 08540     USA}
          \email{witten@sns.ias.edu}   
\begin{abstract}
We analyze the conformal invariance of submanifold observables associated
with $k$-branes in the AdS/CFT correspondence.  For odd $k$, the resulting
observables are conformally invariant, and for even $k$, they transform
with a conformal anomaly that is given by a local expression which we analyze
in detail for $k=2$.
\end{abstract}
 
\maketitle

\thispagestyle{empty}

\renewcommand{\thefootnote}{}
\footnotetext{This research was partially supported by NSF Grant
PHY-9513835.} 

\section{Introduction}\label{intro}
\def\bar{\overline}

There has been much  recent interest in  the correspondence
\cite{maldacena} between  conformal field theory 
and string theory on negatively curved or Anti de Sitter spacetimes.
In this correspondence, correlation functions of local conformal
fields on an $n$-manifold $M$ are computed in terms of the asymptotic
behavior of fields on $n+1$-dimensional Einstein manifolds $X$
of conformal boundary $M$ 
\cite{gubser},\cite{witten}.  (Here
we are actually suppressing the role of some compactified dimensions that
will be unimportant for the present paper.  
The string theory is actually formulated on $X\times W$ for a suitable
compact manifold $W$; the dimension of $X\times W$ is 10 or 11 depending
on whether we are doing string theory or $M$-theory. We will generally
suppress mention of $W$.)

The relevant notion of conformal boundary is that of conformal infinity in
the sense introduced by Penrose (see \cite{penrose}). 
$X$ is the interior of an $n+1$-dimensional manifold-with-boundary
$\overline X$ whose boundary is $M$.  The metric $g_+$ on $X$ is complete
and has a double pole on the boundary in the following sense.  
If $r$ is a smooth function on $\overline X$
with a first order zero on the boundary of $\overline X$,  and
positive on $X$, then $r$ is called a defining function.
The requirement on $g_+$ is that for any defining function $r$,
$\bar g=r^2g_+$ extends as a smooth metric
on $\bar X$.  
Clearly, if so, the restriction of $\bar g$ to $M$ gives a metric on $M$.
This metric changes by a conformal transformation if the defining function
is changed, so $M$ has a well-defined conformal structure but not a 
well-defined
metric.  If $X$ is an Einstein manifold, then 
with a suitable choice of coordinates the metric of $X$ looks like
\begin{equation}\label{choicecord} g_+=\frac{1}{r^2}\left(g_r+dr^2\right),
\end{equation}
with $g_r$ an $r$-dependent family of metrics on $M$; the metric
on $M$ is just $g_0$.


The converse problem -- given a conformal structure on $M$, find an
Einstein metric on $X$ which induces it at infinity -- also has a solution
under certain conditions.  For example, if $M={\bf S}^n$, endowed
with a conformal structure sufficiently close to the standard one, and
$X$ is an $n+1$-ball, then \cite{grahamlee}
 there is a unique Einstein
metric on $X$ with a prescribed negative curvature that is close to the
standard hyperbolic metric on the ball and has $M$ as conformal infinity.

In relating conformal field theory on $M$ to quantum gravity on $X$,
ultraviolet divergences on $M$ are typically related to infrared divergences
on $X$.  This is briefly described in \cite{witten};  the relation of the
statement to ideas about holography is explained somewhat more
fully in \cite{susskindwitten}.  
For example, to compute the partition function
of the conformal field theory on $M$, as a function of the metric of
$M$, one must, in the supergravity approximation,
 evaluate the gravitational action for an Einstein metric on $X$
that induces at infinity the given conformal structure on $M$.
   For an Einstein metric, the gravitational action
is proportional to the volume.  Thus, to compute the partition function
of the conformal field theory, one formally must evaluate the volume
\begin{equation}\label{aaa}{\rm Vol}(X)=\int_Xdv_X,\end{equation} 
where $dv_X=\sqrt {\det{g_+}} d^{n+1}x$ is the Riemannian volume form of $X$.
This integral clearly diverges in view of the form (\ref{choicecord})
of the metric on $X$.  One regularizes it by letting $X_\epsilon$
be the subset of $X$ with $r\geq \epsilon$, and 
\begin{equation}\label{renver}{\rm Vol}_\epsilon(X)=
\int_{X_\epsilon}dv_X.\end{equation}
${\rm Vol}_\epsilon(X)$ has terms that diverge as negative
powers of $\epsilon$,
and also a logarithmic term if $n$ is even.  After subtracting the divergent
terms, one gets a renormalized functional ${\rm Vol}_R(X)$, which,
for a suitable choice of $X$, determines the supergravity limit of the
conformal field theory partition function on $M$.  As shown in
detail in \cite{henningsonskenderis}, the logarithmic term in
${\rm Vol}_\epsilon(X)$ leads to a conformal anomaly in ${\rm Vol}_R(X)$
that reproduces the expected conformal anomaly of the conformal field
theory on $M$. 

The conformal anomaly is possible because the regularization violates
conformal invariance.  Indeed, the choice of a particular defining
function $r$ is built into the definition of $X_\epsilon$.  But the
choice of $r$ violates conformal invariance, because it fixes a particular
metric (namely $r^2g_+$) 
 in the conformal class of metrics on $M$.  The anomaly means that,
 for even $n$, conformal invariance is not restored after renormalization.
 
 \bigskip\noindent{\it Submanifolds}

Our goal in the present paper is to analyze a somewhat similar problem
concerning submanifolds of $X$.  
Consider, for example, Type IIB superstring theory on $X\times {\bf S}^5$,
where $X$ is an Einstein manifold (of negative scalar curvature) with
conformal boundary a four-manifold $M$. 
It has been argued \cite{newmaldacena},\cite{reyyee} 
that to compute the expectation value  of a Wilson line
on a circle $N\subset M$, one must consider a path integral with
a string whose worldsheet $Y\subset \overline X$ has $N$ as boundary.
\footnote{$Y$ is actually a submanifold of $\overline X\times W$.
For simplicity, we will consider only the case that $Y$ projects
to a point in $W$ and so can be regarded as a submanifold of $\overline X$.}
In the supergravity approximation,
$Y$ should obey the equation of a minimal area surface.
Formally speaking, if $A(Y)$ is the area of $Y$ and $T$ the 
string
tension, then the expectation
value of the Wilson line on $N$ is
\begin{equation}
\label{wilsonform}\langle W(N)\rangle =\exp(-TA(Y))\end{equation}
in the supergravity approximation.  Here $W(N)$ is the Wilson line operator
and $\langle W(N)\rangle$ is its expectation value.

Similarly, one can consider examples in which the string theory
contains a $k$-brane for some $k$. Then one associates an observable
$W(N)$ to a $k$-dimensional submanifold $N$ of $M$.
Its expectation value is computed by a path integral with a brane
wrapped on a $k+1$-dimensional submanifold $Y\subset \overline X$
whose boundary is $N$.
In the supergravity approximation, the expectation value of the surface
observable $W(N)$ is again given formally by  (\ref{wilsonform}),
with $A(Y)$ now the volume of $Y$.

The trouble with this formula is that, given the form of the metric 
(\ref{choicecord}), $A(Y)$ is always infinite.  Hence, one proceeds
just as one does in renormalizing  the volume ${\rm Vol}(X)$.  One
lets $Y_\epsilon$ be the part of $Y$ with $r\geq \epsilon$, and one
denotes the volume of $Y_\epsilon$ as $A_\epsilon(Y)$.  As $\epsilon\to 0$,
$A_\epsilon(Y)$ has divergent terms that are negative powers of $\epsilon$,
plus a logarithmic term if $k$ is even.  As explained in 
\cite{witten}, the divergent terms correspond to ultraviolet divergences
in conformal field theory. After subtracting the divergent
terms, one gets a renormalized volume functional $A_R(Y)$ which should
be used instead of $A(Y)$ in (\ref{wilsonform}).  However, when $k$ is even,
$A(Y)$ is not conformally invariant.  The violation of conformal
invariance is given by a local expression that we will analyze.
The anomaly is possible, of course, because the definition of $Y_\epsilon$
depends on the choice of the defining function $r$ and so violates conformal
invariance.

Note that the problem of defining the volume of a submanifold is in a 
very precise sense a generalization of the problem of defining ${\rm Vol}(X)$.
Indeed,    in the special case $k=n$, $N=M$, $Y=X$, $A_R(Y)$
coincides with ${\rm Vol}_R(X)$.

Questions of existence and regularity of minimal area submanifolds (usually
of hyperbolic space) with prescribed boundary at infinity have been studied
in the mathematical literature; see \cite{anderson1}, \cite{anderson2},
\cite{hardtlin}, \cite{lin1}, \cite{lin2}, \cite{tonegawa}.  (An error in 
\cite{lin1} is corrected in \cite{tonegawa}.)

\bigskip\noindent{\it Examples}

We conclude this introduction with comments on a few of the basic
examples to which the discussion can be applied.

First we consider examples with zerobranes, that is with $k=0$.
For example, in Type IIB on $X\times {\bf S}^3\times {\bf T}^4$, with
$X$ a hyperbolic three-manifold, zerobranes on $X$ arise from onebranes
wrapped on a one-cycle in ${\bf T}^4$.  Let $\mu$ be the mass of the zerobrane
(in units in which the Einstein equations on $X$ read $R_{ij}=-3g_{ij}$).
In the regime in which supergravity formulas such as (\ref{wilsonform})
are valid, one has $\mu>>1$.   The conformal boundary of $X$ is a Riemann
surface $M$.

Since $k=0$, the brane world-volume is a curve $Y\subset X$; its boundary
consists of a pair of points $P,Q\in M$.  The operator associated with the
endpoint of the brane worldvolume is thus a local operator ${ \Psi}(P)$.
Actually, assuming that the zerobrane worldvolume is oriented
(as in the example noted in the last paragraph), the points $P,Q$
are endowed with opposite orientations and conjugate operators
${\Psi}(P)$, $\overline{\Psi}(Q)$.  To compute the two point function
$\langle {\Psi}(P)\overline {\Psi}(Q)\rangle$ in the supergravity
approximation, one takes $Y$ to be a minimum length geodesic connecting
$P$ and $Q$.  One then has asymptotically
\begin{equation}\label{length}
\langle {\Psi}(P)\overline {\Psi}(Q)\rangle =\exp(-\mu L(Y)),
\end{equation}
with $L(Y)$ the ``length'' of $Y$.

Here, because $k=0$ is even, we encounter a conformal anomaly.  In fact,
it is clear without any analysis at all that there must be an anomaly.
If  the correlation function (\ref{length}) 
did not require some renormalization leading to an anomaly, then
this correlation function would be conformally invariant. Hence
 the operators ${\Psi}$, $\overline
{\Psi}$ would have conformal dimension zero, a behavior which is impossible
(for non-constant local operators)
in a unitary conformal field theory.

The actual computation of the anomaly is straightforward for $k=0$.
The  length is regularized by replacing $L(Y)$ by $L_\epsilon(Y)$, the
length of the part of $Y$ with $r\geq \epsilon$.  
As $\epsilon\to 0$, $L_\epsilon(Y)$ receives a divergent contribution
from each of the two ends of $Y$.
With the metric in (\ref{choicecord}),
the divergent contribution of either end is precisely $+1\cdot \ln 1/\epsilon$.
Hence the two point function in (\ref{length}), after regularization,
has the form
\begin{equation}\label{beheq}
\langle{\Psi}(P)\overline{\Psi}(Q)\rangle_\epsilon = 
\epsilon^{2\mu}S(\epsilon),
  \end{equation}
  where  $ S(\epsilon)$ has a limit as $\epsilon\to 0$.  One defines the
  renormalized two point function $\langle \Psi(P)\overline\Psi(Q)\rangle_R$
  to be the limit $S(0)$.  
  Since $\epsilon$ has dimensions
  of length or $({\rm mass})^{-1}$, 
  $S$ has dimensions of $({\rm mass})^{2\mu}$, and hence
  $\Psi$ and $\bar \Psi$ have conformal dimension $\mu$.
By contrast, in the general AdS/CFT correspondence, an AdS particle
of mass $\mu$ should (in $n$ dimensions)
correspond to a conformal field of dimension
$d(\mu)=(n+\sqrt{n^2+4\mu^2})/2$.  This reduces to $d(\mu)=\mu$ in the large
$\mu$ limit, where (\ref{length}) is valid, so the anomaly reproduces
the expected conformal dimension of ${ \Psi}$ and $\overline{ \Psi}$ in
this limit.

The next case is $k=1$.  The most familiar example is that of
't Hooft and Wilson loops in four-dimensional gauge theory.  There is
no anomaly for odd $k$, so these operators have
no conformal anomaly. 

The next case -- and the last one that we will consider specially -- 
is $k=2$.  Examples are known for various values of $n$ up to $n=6$.
For instance, $k=2$, $n=6$ arises in the case of
$M$-theory on $AdS_7\times {\bf S}^4$.  By replacing ${\bf S}^4$ with
a quotient, one can also build other $k=2$, $n=6$
examples with less supersymmetry.
For $k=2$ and $n=6$, the operators $W(N)$ are ``surface'' observables
in a six-dimensional conformal field theory.  As $k=2$ is even, an anomaly
occurs; it will be analyzed in detail in the next section.
Because of our limited understanding of the six-dimensional conformal
field theories, no independent way to compute this anomaly is presently known,
so  there is no theory for us to compare to.

The computation of the anomaly for $k=2$ thus
gives new information about the conformal field theories that arise
from these constructions.   
In section 2 of this paper, we will define the anomaly precisely,
and describe some of its general properties, for all $k$.  Then
we will do a detailed computation for $k=2$.

The anomaly for $k=2$ has also been briefly
discussed,
in the case that $M$ is conformally flat, in \cite{berenstein}, which
appeared while the present paper was in gestation.

\section{The Computation}\label{compu}

Let $X^{n+1}$ be the interior of a compact manifold with boundary
$\Xc$, let $M=\partial X$, 
and let $g_+$ be a conformally compact metric on $X$.  This means that if
$r$ is a defining function for $M\subset \Xc$ (in a sense explained in the
introduction), then $\gb=r^2 g_+$
extends smoothly to $\Xc$.  The conformal class of the restriction of $\gb$
to $TM$ is independent of the choice of defining function.
 The function $|dr|^2_{\gb}$ extends
smoothly to 
$\Xc$, and its restriction to $M$ is independent of the choice
of $r$, so is an invariant of $g_+$.  We will assume that $g_+$
satisfies the Einstein condition $Ric(g_+)= -ng_+$.  
It follows upon conformally transforming the Einstein equation that 
in this case one has $|dr|^2_{\gb} =1$ on $M$.  

Conformally compact 
metrics with $|dr|^2_{\gb} =1$ on $M$ 
may be put in a special form near the boundary using a
special class of defining functions.  The following Lemma is taken from
\cite{grahamlee} (see Lemma 5.2).

\begin{lemma}\label{deffn}
A metric on $M$ in the conformal
infinity of $g_+$ determines a unique defining function $r$ in 
a neighborhood of $M$ in $\Xc$ such that $\gb|_{TM}$ is the prescribed
boundary metric and such that $|dr|^2_{\gb}=1$.
\end{lemma}
\begin{proof}
Given any choice of defining function $r_0$, 
let $\gb_0=r^2_0 g_+$ and set $r=r_0 e^{\om}$, so $\gb=e^{2\om}\gb_0$
and $dr=e^\om (dr_0+r_0 d\om )$.  Thus
\begin{equation}\label{firsteq}
|dr|^2_{\gb}=|dr_0 + r_0d\om |^2_{\gb_0}=
|dr_0|^2+2r_0(dr_0,d\om )+r_0^2|d\om |^2_{\gb_0}\end{equation}
(where $(~,~)$ is the inner product in the metric $g_0$),
so the condition $|dr|^2_{\gb}=1$ is equivalent to
\begin{equation}\label{pde}
2(dr_0,d\om )+r_0|d\om |^2_{\gb_0}=
\frac{1-|dr_0|^2_{\gb_0}}{r_0}.
\end{equation}
This is a non-characteristic first order PDE for $\om$, so there is a
solution near $M$ with $\om |_{M}$ arbitrarily prescribed.
\end{proof}

This lemma means that not only does a defining function $r$ determine
a metric on $M$ in its conformal class, but conversely given such a metric
on $M$, there is a natural way to determine a distinguished defining function
$r$, at least in a neighborhood of $M$.   Since we will only be interested
in the behavior of $r$ near $M$, it follows that for our purposes, the choice
of a metric on $M$ in its conformal class is equivalent to the choice of
a defining function.  

A defining function determines for some $\epsilon>0$ an
identification of $M \times [0,\epsilon)$ with a neighborhood of
$M$ in $\Xc$: $(p,\lambda)\in M\times[0,\epsilon)$
corresponds to the point obtained by following the integral curve of
$\nabla_{\gb}r$ emanating from $p$ for $\lambda$ units of time.  For a defining
function of the type given in the lemma, with
$|dr|^2_{\gb}=1$, the $\lambda$-coordinate is just $r$, and
$\nabla_{\gb}r$ is orthogonal to the slices
$M\times\{\lambda\}$.  Hence, identifying $\lambda$ with $r$,
on $M\times[0,\epsilon)$ the metric $\gb$ takes the form
$\gb=g_{r}+dr^2$ for a 1-parameter family $g_{r}$ of metrics on
$M$, and 
\begin{equation}\label{form}
g_+=r^{-2}(g_{r}+dr^2).
\end{equation}

One can explicitly calculate the Ricci curvature of a metric of the form
(\ref{form}) so as to express the Einstein condition directly in terms of
$g_r$.  From an analysis of the formal asymptotics of solutions of the
resulting equations (see \cite{feffermangraham} or 
\cite{henningsonskenderis}), 
one deduces that for $n$ odd, the expansion of $g_r$ is of the form
\begin{equation}\label{oddasym}
g_r= g^{(0)} + g^{(2)} r^2 + (even\, powers) + g^{(n-1)} r^{n-1} +
g^{(n)} r^n + \ldots,
\end{equation}
where the $g^{(j)}$ are tensors on $M$, and $g^{(n)}$ is trace-free with
respect to a metric in the conformal class on $M$.  For $j$ even
and $0\leq j \leq n-1$, the tensor $g^{(j)}$ is locally formally determined by 
the
conformal representative, but $g^{(n)}$ is formally
undetermined, subject to the trace-free condition.  For $n$ even the
analogous expansion is
\begin{equation}\label{evenasym}
g_r= g^{(0)} + g^{(2)} r^2 + (even\, powers) + kr^n \log r +
g^{(n)} r^n + \ldots,
\end{equation}
where now the $g^{(j)}$ are locally determined for $j$ even and $0\leq j
\leq n-2$, $k$ is locally determined and trace-free, the trace of
$g^{(n)}$ is locally determined, but the trace-free part of $g^{(n)}$ is
formally undetermined.  Moreover, so long as $n\geq 3$, one has
\begin{equation}\label{P}
g^{(2)}_{ij} = - P_{ij},
\end{equation}
where  
$(n-2)P_{ij} = R_{ij} - \frac{R}{2(n-1)}g_{ij},$ and $R_{ij}$ and $R$
denote 
the Ricci tensor and scalar curvature of the chosen representative $g_{ij}$
of the conformal infinity.

Later we will need to use the following Lemma.
\begin{lemma}\label{rel}
Let $r$ and $\rh$ be special defining functions as in Lemma~\ref{deffn}
associated to two different conformal representatives.  Then 
\begin{equation}\label{relation}
\rh = r e^{\om}
\end{equation} 
for a function $\om$ on $M\times[0,\ep)$ whose expansion at
$r=0$ consists only of even powers of $r$ up through and including the
$r^{n+1}$ term.
\end{lemma}
\begin{proof}
We have $\rh=e^{\om}r$ where $\om$ is determined by (\ref{pde}), which in
this case becomes 
\begin{equation}\label{newpde}
2\om_r +r(\om_r^2+|d_M\om|_{g_r}^2)=0.
\end{equation}  
The Taylor expansion of $\om$ is determined inductively by differentiating
this equation at $r=0$.  Clearly $\om_{r} = 0$ at $r=0$.  Consider
the determination of $\partial_r^{k+1}\om$ resulting from
differentiating (\ref{newpde}) an even number $k$ times and
setting $r=0$.  The term $\om_r^2$ gets differentiated $k-1$ times, 
so one of the two factors ends up differentiated an odd number of
times, so by induction vanishes at $r=0$.  Now
$|d_M\om|_{g_r}^2=g_{r}^{ij}\om_i\om_j$, so the $k-1$ differentiations
must be split between the three factors, so one of the factors 
must receive an odd
number of differentiations.  When an odd number of derivatives hits a $\om_i$,
the result again vanishes by induction.  But by (\ref{oddasym}) and
(\ref{evenasym}), so long as $k-1<n$, the odd derivatives of $g_r$ vanish
at $r=0$.  
\end{proof}

We will calculate near $M$ the minimal surface equation for a submanifold
of $(X,g_+)$.  We begin by deriving the minimal surface equation
for a graph in a Riemannian manifold. 
 
Let
$(x^{\al},u^{\al'})$ denote coordinates in $\mathbb{R}^m \times
\mathbb{R}^l$, let $g$ be a metric on $\mathbb{R}^m \times \mathbb{R}^l$,
and let $Y$ denote the graph $\{u=u(x)\}$.  Then $g$ restricts to $Y$ to
the metric $h$ given in the coordinates $x$ on $Y$ by
\begin{equation}\label{indmetric}
h_{\al\be}=g_{\al\be} + 2g_{\al'(\al}u^{\al'}_{,\be)} +
g_{\al'\be'}u^{\al'}_{,\al}u^{\be'}_{,\be},
\end{equation}
where the indices after a comma indicate coordinate differentiation. 
The area of $Y$ is $A=\int \sqrt{\det h} \, dx$, so 
$
\delta A = \frac{1}{2} \int \sqrt{\det h} \, h^{\al\be} 
\delta h_{\al\be} \,dx.
$
Since 
$$
\delta h_{\al\be} = \left [g_{\al\be,\ga'} +
2g_{\al'(\al,|\ga'|}u^{\al'}_{,\be)} 
+ g_{\al'\be',\ga'}u^{\al'}_{,\al}u^{\be'}_{,\be}\right ] \delta u^{\ga'} +  
2g_{\al'(\al} \delta u^{\al'}_{,\be)} + 2 g_{\al'\be'} u^{\al'}_{,(\al}
\delta u^{\be'}_{,\be)},
$$
it follows that $Y$ is stationary for area iff $u$ satisfies
$$
2\left [ \sqrt{\det h} \, h^{\al\be} \left ( g_{\al\ga'}
+ g_{\al'\ga'}u^{\al'}_{,\al} \right ) \right ]_{,\be} - \sqrt {\det h} \,
h^{\al\be} \left [ g_{\al\be,\ga'} +  2g_{\al\al',\ga'}u^{\al'}_{,\be}
+ g_{\al'\be',\ga'}u^{\al'}_{,\al}u^{\be'}_{,\be} \right ] = 0,
$$
which may be rewritten as
\begin{multline}\label{minsurf}
\left [ \partial _{\be} +  \frac{1}{2} \left ( \log (\det h) 
\right )_{,\be} \right ] 
\left [ h^{\al\be} \left ( g_{\al\ga'} + g_{\al'\ga'}u^{\al'}_{,\al} 
\right ) \right ] \\
 - \frac{1}{2} h^{\al\be}\left [ g_{\al\be,\ga'} +
2g_{\al\al',\ga'}u^{\al'}_{,\be} 
+ g_{\al'\be',\ga'}u^{\al'}_{,\al}u^{\be'}_{,\be} \right ] = 0.
\end{multline}

Let now $Y^{k+1}$ be a submanifold of our conformally compact Einstein 
manifold 
$(X,g_+)$, where $0\leq k \leq n-1$.  
Suppose that $Y$ extends smoothly to $\overline{X}$ and set $N = Y\cap
M$.   Locally near a point of $N$, coordinates
$(x^{\al},u^{\al'})$ for $M$ may be chosen, where $1\leq \al \leq k$ and
$1\leq \al' \leq n-k$, so that $N=\{u=0\}$ and so that $\partial_{x^{\al}}
\perp \partial_{u^{\al'}}$ on $N$ with respect to 
a metric in the conformal infinity of $g_+$. 
Under the identification discussed above, the choice of such a 
metric
determines an extension of the $x^{\al}$ and $u^{\al'}$ to a neighborhood
of $M$, and together with $r$ these form a coordinate system on $\Xc$.
We consider submanifolds $Y$ which in these coordinates may be written as a 
graph $\{u=u(x,r)\}$.  The minimal surface equation for $Y$ is therefore
given by (\ref{minsurf}), where however $x^{\al}$ must be replaced by
$(x^{\al},r)$ and $g$ by $g_+$.  Set $\hb=r^2 h$ and 
$L = \log (\det \hb)$. Recalling 
(\ref{form}) and writing simply $g$ for $g_{r}$, 
the minimal surface equation for $Y$ becomes $\M (u) = 0$, where
\begin{equation}\label{ccminsurf}
\begin{aligned}
\M (u)_{\ga'} = &
\left [ r \partial_{r} - (k+1) + \frac{1}{2} rL_{,r} \right ] 
\left [ \hb^{rr} g_{\al'\ga'}u^{\al'}_{,r} +
\hb^{\al r} \left ( g_{\al\ga'} + g_{\al'\ga'}u^{\al'}_{,\al} 
\right ) \right ]\\
& +r\left [ \partial_{\be} + \frac{1}{2} L_{,\be} \right ]
\left [ \hb^{r\be} g_{\al'\ga'}u^{\al'}_{,r} +
\hb^{\al\be} \left ( g_{\al\ga'} + g_{\al'\ga'}u^{\al'}_{,\al} 
\right ) \right ] \\
& -\frac{1}{2} r \hb^{\al\be}\left [ g_{\al\be,\ga'} +
2g_{\al\al',\ga'}u^{\al'}_{,\be} 
+ g_{\al'\be',\ga'}u^{\al'}_{,\al}u^{\be'}_{,\be} \right ] \\
& -r \hb^{\al r}\left [ g_{\al\al',\ga'}u^{\al'}_{,r} 
+ g_{\al'\be',\ga'}u^{\al'}_{,\al}u^{\be'}_{,r} \right ] \\
& -\frac{1}{2} r \hb^{rr}\left [ 
g_{\al'\be',\ga'}u^{\al'}_{,r}u^{\be'}_{,r} \right ].
\end{aligned}
\end{equation}

Consider now the inductive determination of the expansion of $u=u(x,r)$ 
from the equation $\M(u)=0$,
beginning with the initial condition $u(x,0)=0$.  Since the coordinates
were chosen so that $g_{\al\ga'}=0$ on $N$, the representative $g_{ij}$ for
the conformal structure decomposes at $u=r=0$ into two pieces 
$g_{\al\be}$ and $g_{\al'\be'}$.  Upon setting $r=0$ and using
the initial condition, all terms on the right hand side of (\ref{ccminsurf})
vanish except for the first, and one obtains $u_r=0$ 
at $r=0$.  Thus a minimal $Y$ must intersect the boundary
orthogonally.
Using this, one deduces that all terms on the right hand side of 
(\ref{ccminsurf}) are $O(r^2)$ except the first and third, so that
$\left ( r \partial_{r} - (k+1) \right ) \left ( \hb^{rr}
g_{\al'\ga'}u^{\al'}_{,r} \right ) - \frac{1}{2} r \hb^{\al\be}
g_{\al\be,\ga'} =O(r^2).$  This gives 
$ku^{\ga'}_{rr} = -\frac{1}{2} g^{\al'\ga'}g^{\al\be}g_{\al\be,\al'}$ at
$r=0$.  Of course when $k=0$ there are no unprimed indices so that this
equation reads $0=0$, recovering the fact familiar from hyperbolic geometry
that at the boundary a geodesic may have any asymptotic curvature measured
using the smooth metric $\gb$.

Recall that the second fundamental form $B^{\ga'}_{\al\be}$ of $N$ with
respect to 
$g_{ij}$ is defined by $B(X,Y) = (\nabla_X Y)^{\perp}$ for
vectors $X,Y \in TN$; here $\nabla$ denotes the Levi-Civita covariant
derivative of $g_{ij}$ and $\perp$ the component in $TN^{\perp}$.  
$B$ is symmetric in its lower indices.  The mean
curvature vector of $N$ is $H^{\ga'} = g^{\al\be} B^{\ga'}_{\al\be}$.
In our coordinates $(x,u)$ on $M$, the definition immediately gives 
$B^{\ga'}_{\al\be}=\Gamma^{\ga'}_{\al\be}$, where $\Gamma$ denotes the
Christoffel symbol.  Since $g_{\al\ga'}=0$ on $N$, this gives 
\begin{equation}\label{22form}
B^{\ga'}_{\al\be}=-\frac{1}{2}g^{\al'\ga'} g_{\al\be,\al'},
\end{equation} 
so that the
above formula for $u^{\ga'}_{rr}$ at $r=0$ becomes 
\begin{equation}\label{2deriv}
ku^{\ga'}_{rr} = H^{\ga'}.
\end{equation}

Inductively, suppose that $v$ has been determined so that $\M(v) =
O(r^{m-1})$.  Set $u=v+wr^{m}$, where $w$ is to be determined.  Then it is 
not hard to see that (supressing the raising and lowering of indices)
$\M(u)=\M(v) +m(m-k-2)wr^{m-1} + O(r^m).$ 
It follows that so long as
$m<k+2$ then $w$ is uniquely determined.  When $m=k+2$, $w$ is
formally undetermined and one must include a 
$r^{k+2} \log r$ term in $u$ if the corresponding $\M(v)$ is not actually
already  $O(r^{k+2})$.

Since by (\ref{oddasym}) and (\ref{evenasym}) the metric $g_{ij}$ is even
in $r$ to high order, it follows by staring at (\ref{ccminsurf}) that the
expansion determined for $u$ will also be even in $r$ up to the $r^{k+2}$
term.  (Alternatively, one may introduce $r^{\frac{1}{2}}$ as
a new variable and observe that (\ref{ccminsurf}) 
remains regular.)  When $k$ is odd, the same reasoning shows that the 
$r^{k+2} \log r$ term in $u$ mentioned above does not occur, but instead
the parity is broken at this point and $u$ may have a formally undetermined 
$r^{k+2}$ term.  Thus we have for $k$ odd
\begin{equation}\label{uodd}
u= u^{(2)} r^2 + (even\, powers) + u^{(k+1)} r^{k+1} +
u^{(k+2)} r^{k+2} + \ldots,
\end{equation}
and for $k$ even
\begin{equation}\label{ueven}
u= u^{(2)} r^2 + (even\, powers) + u^{(k)} r^{k} +
vr^{k+2} \log r + u^{(k+2)} r^{k+2} + \ldots,
\end{equation}
where the $u^{(j)}$ and $v$ are functions of $x$, all of which are locally 
determined except for $u^{(k+2)}$.

The induced metric $h$ on $Y$ is given by (\ref{indmetric}) with $x^{\al}$
replaced by $(x^{\al},r)$ and $g$ by $g_+$.  Since the irregularities in
$g_r$ occur at order $n$ and those in $u$ at order $k+2$, 
one concludes that up to
terms vanishing to order greater than $k$, the expansions of $\hb_{\al\be}$ 
and $\hb_{rr}$ have only even terms in $r$ and that of $\hb_{\al r}$ has
only odd terms.  Thus the volume form $dv_Y=\sqrt{\det h} \,dxdr$ takes
the form 
\begin{equation}\label{volform}
dv_Y=r^{-k-1}\left [ v^{(0)} + v^{(2)} r^2 + (even\, powers) + v^{(k)}
r^{k} +\ldots \right ] dv_Ndr,
\end{equation}
where the $\ldots$ indicates terms vanishing to higher order and $dv_N$
denotes the volume form on $N$ with respect to the chosen conformal
representative on the boundary.  All indicated $v^{(j)}$ are locally
determined functions on $N$ and $v^{(k)} =0$ if $k$ is odd.  

Consider now the asymptotics of Area$_{g_+}(Y \cap \{r>\ep\})$ as 
$\ep \rightarrow 0.$  Fix a small
number $r_0$ and express
$\mbox{Area}(Y \cap \{r>\ep\})= C + \int_{Y \cap \{\ep<r<r_0\}}dv_Y$.  
By (\ref{volform}) we obtain for $k$ odd
\begin{equation}
\mbox{Area}(Y \cap \{r>\ep\})= c_0 \ep^{-k} + c_2\ep^{-k+2} +
(even\, powers) + c_{k-1} \ep^{-1} + c_k + o(1)
\end{equation}
and for $k$ even
\begin{equation}\label{ceven}
\mbox{Area}(Y \cap \{r>\ep\})= c_0 \ep^{-k} + c_2\ep^{-k+2} +
(even\, powers) + c_{k-2} \ep^{-2} + d\log {\frac{1}{\ep}} +c_k + o(1).
\end{equation}
Observe that 
\begin{equation}\label{dform}
d = \int_N v^{(k)}\, dv_N.
\end{equation}
\begin{proposition}\label{invariance}
If $k$ is odd, then $c_k$ is independent of the special defining function.
If $k$ is even, then $d$ is independent of the special defining function.
\end{proposition}
\begin{proof}
Let $r$ and $\rh$ be two special defining functions.  On $Y$, 
(\ref{relation}) becomes 
\begin{equation}\label{defrel}
\rh = r e^{\om(x,u(x,r), r)}.
\end{equation}  
{}From Lemma \ref{rel} and (\ref{uodd}),
(\ref{ueven}), we see that 
the expansion of $e^{\om(x,u(x,r), r)}$ at $r=0$ has only even powers of
$r$ up through and including the $r^{k+1}$ term.  So (\ref{defrel}) can be
solved for $r$ to give $r=\rh b(x,\rh)$ on $Y$, where the expansion of $b$
also has only even powers of $\rh$ up through the $\rh^{k+1}$ term.  
It is important to note that in this
relation the $x$ still refers to the identification associated with $r$. 

Set $\eh(x, \ep)=\ep b(x,\ep)$.  Then on $Y$,
$\rh>\ep$ is equivalent to $r>\eh(x,\ep)$, so 
$$
\mbox{Area}_{g_+}(Y\cap \{r>\ep\})-\mbox{Area}_{g_+}(Y\cap \{\rh>\ep\})= 
\int_{N}\int_{\ep}^{\eh}dv_Y=
$$
\begin{equation}\label{diff}
\int_N\int_{\ep}^{\eh}
\sum_{\stackrel{0\leq j\leq k}{j \,
even}} v^{(j)}(x) r^{-k-1+j}dr dv_N + o(1),
\end{equation}
where we have used (\ref{volform}).
For $k$ odd this is 
$$
\sum_{\stackrel{0\leq j\leq k-1}{j \,even}}\ep^{-k+j}
\int_N\frac{v^{(j)}(x)}{-k+j}\left ( b(x,\ep)^{-k+j}-1 \right )dv_N +
o(1). 
$$
Since $b(x,\epsilon)$ is even through terms of order $k+1$ in $\ep$,
it follows that this expression has no constant term as
$\epsilon \rightarrow 0$.  Similarly, when $k$ is even, the $r^{-1}$ term
in (\ref{diff}) contributes $\log b(x,\ep)$, so there is no
$\log {\frac{1}{\ep}}$ term as $\epsilon \rightarrow 0$. 
\end{proof}

As explained in the introduction, we define the renormalized volume of $Y$
to be  $A_R(Y)=c_k$.  The lemma says that $A_R(Y)$ is conformally
invariant if $k$ is odd.  This is not so if $k$ is even.
For $k$ even, the variation in $c_k$ under a conformal change is the constant
term in the expansion of (\ref{diff}).  It is clear from the local
determination of the $v^{(j)}$ and of the expansion of $b(x,\ep)$ that this 
anomaly takes the form of the integral over $N$ of a locally determined
expression.  

\bigskip\noindent
{\it The Anomaly For Surface Observables}

As described in the introduction, the anomaly can be extracted
rather trivially for $k=0$.  So the first example that really illustrates
this computation is $k=2$.
In the remainder of the paper, we compute the anomaly as well as the 
log term coefficient $d$ for $k=2$.

Using (\ref{indmetric}) with $g$ again replaced by $g_+$ and $x^{\al}$ by
$(x^{\al},r)$ and recalling $\hb = r^2 h$, one concludes that 
the induced metric on $Y$ satisfies
$$
\begin{aligned}
&\hb_{\al\be} = g_{\al\be} + O(r^3) \\
&\hb_{\al r} = O(r^3) \\
&\hb_{rr} = 1 + g_{\al'\be'} u^{\al'}_{,r} u^{\be'}_{,r}. 
\end{aligned}
$$
If we denote by $P_{\al\be}$ the $TN$-component 
of $P_{ij}$, then (\ref{P}) and (\ref{2deriv}) give for $k=2$
$$
\begin{aligned}
g_{\al\be}(x,u,r)&=g_{\al\be}(x,u,0) - P_{\al\be}(x,u,0)r^2 +O(r^3)\\
&=g_{\al\be}(x,0,0) +g_{\al\be,\ga'}(x,0,0)u^{\ga'}(x,r) 
- P_{\al\be}(x,0,0)r^2 +O(r^3) \\ 
&=g_{\al\be}(x,0,0) - \left (\frac{1}{2}H_{\ga'}B_{\al\be}^{\ga'}
+P_{\al\be} \right )r^2 +O(r^3),
\end{aligned}
$$
and
$$
\begin{aligned}
\hb_{rr}(x,r)&=1+g_{\al'\be'}(x,0,0) u^{\al'}_{,r}(x,r) u^{\be'}_{,r}(x,r)
+O(r^3) \\ 
&=1+\frac{1}{4}|H|^2r^2 +O(r^3).
\end{aligned}
$$
Therefore, if we now use $g_{\al\be}$ to denote the values on
$N=\{u=r=0\}$, we have
$$
\begin{aligned}
\det \hb &= (\det \hb_{\al\be}) \hb_{rr} +O(r^3)\\
&= (\det  g_{\al\be}) \left [1-\left ( \frac{1}{4} |H|^2
+g^{\al\be}P_{\al\be} \right )r^2 \right ] +O(r^3),
\end{aligned}
$$
so that 
\begin{equation}\label{dv}
dv_Y = r^{-3}\left [1- \frac{1}{8} \left (|H|^2
+4g^{\al\be}P_{\al\be} \right ) r^2 +O(r^3)\right ] dv_Ndr .
\end{equation}
By (\ref{dform}) we obtain for the log term coefficient
$$
-8d=\int_N \left (|H|^2 +4g^{\al\be}P_{\al\be} \right ) dv_N.
$$

By Proposition \ref{invariance}, this quantity is an invariant of the
conformal structure on $M$ and the submanifold $N$.  In conformally flat
spaces it is known in the mathematical literature 
as the Willmore functional of $N$ and its conformal invariance is
well-known.  A problem that has received much attention is that of
minimizing the Willmore functional over all embeddings of a 
2-manifold of fixed topological type.  In the physics literature, as
noted in \cite{berenstein}, this functional has been called the rigid string
action \cite{polyakov}.

To calculate the anomaly, let $\gh_{ij}=e^{2\up}g_{ij}$ 
be a second representative for
the conformal infinity, where $\up\in C^{\infty}(M)$.  The associated
defining functions $\rh$ and $r$ are related by $\rh=e^{\om}r$, where $\om$
solves (\ref{newpde}) and $\om = \up$ on $M$.  Differentiation of
(\ref{newpde}) gives $\om_{rr} = -\frac{1}{2}\up_i\up^i$ at $r=0$,
so 
$$
\begin{aligned}
\om(x,u,r)&=\up(x,u)-\frac{1}{4}\up_i\up^i r^2 +O(r^3) \\
&=\up(x,0)+\up_{\ga'}(x,0)u^{\ga'}(x,r)-\frac{1}{4}\up_i\up^i r^2 +O(r^3) \\
&=\up(x,0)+\frac{1}{4}\left ( \up_{\ga'}H^{\ga'}-\up_i\up^i \right )r^2 
+O(r^3).
\end{aligned}
$$
Solving the equation $\rh=e^{\om(x,u(x,r),r)}r$ for $r$ gives
$r=\rh b(x,\rh)$, where $b=e^{-\Om}$ and
$$
\Om(x,\rh)=\up(x,0) + \frac{1}{4} e^{-2\up}(\up_{\ga'}H^{\ga'}-
\up_i\up^i)\rh^2 +O(\rh^3).
$$
As previously observed, if $c_2$ and $\ch_2$ denote the constant terms in
(\ref{ceven}), then $c_2-\ch_2$ is the constant term in
(\ref{diff}).  By (\ref{dv}), this is the same as the constant term in 
$$
\int_N \left [- \frac{1}{2} ( \ep b(x,\ep))^{-2} -
\frac{1}{8}(|H|^2+4g^{\al\be}P_{\al\be})\log {b(x,\ep)} \right ] dv_N,
$$
which is easily calculated to give
\begin{proposition}
When $k=2$, the anomaly is given by
$$
8(c_2-\ch_2)=\int_N \left [ \up (|H|^2+4g^{\al\be}P_{\al\be})
-2\up_{\ga'}H^{\ga'}+2\up_i\up^i \right ] dv_N.
$$
\end{proposition}

It is interesting to observe that the linearization in $\up$ of this
anomaly involves derivatives of $\up$---terms which do not arise from the 
log term in (\ref{ceven}) upon rescaling $\ep$.

\bigskip

We would like to thank Charles Fefferman and Tatiana Toro for helpful
discussions.

\end{document}